\documentclass[12pt]{article}
\usepackage{amsmath}
\usepackage{graphicx}
\usepackage{natbib}
\usepackage{geometry}
\usepackage{setspace}

\title{Advances in Machine Learning, Statistical Methods, and AI for Single-Cell RNA Annotation Using Raw Count Matrices in scRNA-seq Data}
\author{
    Megha Patel\textsuperscript{1,2}, 
    Nimish Magre\textsuperscript{1}, 
    Himanshi Motwani\textsuperscript{1}, 
    and Nik Bear Brown\textsuperscript{1,2} \\
    \\
    \textsuperscript{1}Northeastern University \\
    \textsuperscript{2}Bear Brown \& Company
}
\date{}

\date{}

\begin{document}
\maketitle

\begin{abstract}
Single-cell RNA sequencing (scRNA-seq) has revolutionized our ability to analyze gene expression at the resolution of individual cells, providing unprecedented insights into cellular heterogeneity and complex biological systems. This paper reviews various advanced computational and machine learning techniques tailored for the analysis of scRNA-seq data, emphasizing their roles in different stages of the data processing pipeline.

We explore multiple machine learning techniques, including dimensionality reduction methods such as Principal Component Analysis (PCA), t-Distributed Stochastic Neighbor Embedding (t-SNE), and Uniform Manifold Approximation and Projection (UMAP), which are crucial for visualizing high-dimensional data and retaining its intrinsic structure. Clustering techniques, including k-means, hierarchical clustering, and graph-based clustering, are reviewed for their efficacy in identifying distinct cell populations based on gene expression profiles.

Classification methods like Support Vector Machines (SVM), Random Forests, and Neural Networks are examined for their ability to accurately categorize cell types, leveraging both supervised and unsupervised learning paradigms. We also discuss the application of statistical techniques, such as normalization (Log Normalization and Scaling), differential expression analysis (Wilcoxon Rank-Sum Test and Likelihood Ratio Test), and batch effect correction (ComBat and Harmony), to enhance data quality and interpretability.

Advanced AI techniques, including Autoencoders, Graph Neural Networks (GNNs), and Generative Adversarial Networks (GANs), are highlighted for their potential to improve feature extraction, clustering accuracy, and synthetic data generation. We also delve into data integration and annotation strategies, such as transfer learning, ensemble methods, and tools like SingleR and SCINA, which enhance the accuracy and robustness of cell type identification.

The paper further outlines a comprehensive data processing pipeline, detailing steps from preprocessing (quality control and handling missing values), feature selection (identifying highly variable genes), and model training (supervised and unsupervised learning) to evaluation (using metrics like accuracy, precision, recall, and F1-score). This pipeline ensures effective analysis of scRNA-seq data, enabling researchers to uncover cellular heterogeneity, identify distinct cell types, and gain insights into biological processes.

By integrating these advanced techniques, we provide a detailed framework for scRNA-seq data analysis, showcasing the interplay of various computational methods in enhancing the understanding of complex biological systems.
\end{abstract}

\section{Introduction}

Single-cell RNA sequencing (scRNA-seq) technology has revolutionized the study of cellular heterogeneity and gene expression at the individual cell level, offering unprecedented insights into complex biological systems. Despite its transformative potential, the analysis of scRNA-seq data poses significant challenges due to its high-dimensional, noisy, and sparse nature. This survey paper reviews various machine learning, statistical, and artificial intelligence (AI) techniques employed for single-cell RNA annotation using raw count matrices from scRNA-seq data.

Key machine learning methodologies explored include dimensionality reduction techniques such as Principal Component Analysis (PCA), t-Distributed Stochastic Neighbor Embedding (t-SNE), and Uniform Manifold Approximation and Projection (UMAP). Clustering methods such as k-means, hierarchical clustering, and graph-based clustering, alongside classification approaches including Support Vector Machines (SVM), Random Forests, and neural networks, are discussed in detail. Additionally, the review delves into statistical methods for data normalization, differential expression analysis, and batch effect correction, emphasizing tools like log normalization, scaling, ComBat, and Harmony.

Key quotes from the literature underscore the advancements and challenges in this field:
\begin{itemize}
    \item "Single-cell RNA sequencing (scRNA-seq) technologies have enabled the large-scale whole-transcriptome profiling of each individual single cell in a cell population."
    \item "A core analysis of the scRNA-seq transcriptome profiles is to cluster the single cells to reveal cell subtypes and infer cell lineages based on the relations among the cells."
    \item "The review focuses on how conventional clustering techniques such as hierarchical clustering, graph-based clustering, mixture models, k-means, ensemble learning, neural networks, and density-based clustering are modified or customized to tackle the unique challenges in scRNA-seq data analysis."
    \item "We review how cell-specific normalization, the imputation of dropouts, and dimension reduction methods can be applied with new statistical or optimization strategies to improve the clustering of single cells."
    \item "Several software packages developed to support the cluster analysis of scRNA-seq data are also reviewed and experimentally compared to evaluate their performance and efficiency."
\end{itemize}

Advanced methods discussed in the paper include:
\begin{itemize}
    \item \textbf{Training Distribution Matching (TDM)}: This method normalizes RNA-seq data for use with models constructed from legacy platforms, demonstrating consistently strong performance in addressing dataset shifts.
    \item \textbf{scGMAI}: A Gaussian mixture clustering method based on autoencoder networks and FastICA, outperforming existing tools like Seurat in clustering accuracy.
    \item \textbf{DRjCC}: A joint learning algorithm for dimension reduction and cell clustering, showing significant improvements in cell type discovery.
    \item \textbf{DESC}: An unsupervised deep embedding algorithm for single-cell clustering that iteratively learns cluster-specific gene expression signatures and removes batch effects while preserving biological variations.
\end{itemize}

Key methods and their contributions:
\begin{itemize}
    \item \textbf{t-SNE and UMAP}: Evaluated for their performance in dimensionality reduction, with t-SNE providing the highest accuracy and computing cost, while UMAP offers the highest stability.
    \item \textbf{NMF}: Demonstrated effectiveness in clustering cells and identifying important genes across various scRNA-seq datasets.
    \item \textbf{SIMLR}: A similarity-learning framework that improves clustering sensitivity, accuracy, and visualization by learning an appropriate distance metric from the data.
    \item \textbf{Consensus Clustering}: Utilizes resampling techniques to assess cluster stability and sensitivity to initial conditions, providing a robust methodology for cluster validation.
    \item \textbf{GOAE and GONN}: Integrate Gene Ontology with neural networks for clustering scRNA-seq data, outperforming existing methods in dimensionality reduction and biological interpretability.
\end{itemize}

The paper emphasizes the need for methods tailored to the unique characteristics of scRNA-seq data, such as high dropout rates and large numbers of cells, and highlights the importance of addressing batch effects and ensuring accurate cell type identification.

In conclusion, this paper provides a comprehensive resource for researchers aiming to leverage machine learning, statistics, and AI in the annotation of single-cell RNA data. It addresses the unique challenges posed by scRNA-seq datasets and illustrates the efficacy of these techniques through practical applications. Further analysis and results will be discussed in subsequent sections to provide deeper insights into the methodologies and their impact on scRNA-seq data analysis.

\section{Machine Learning Techniques}
\subsection{Dimensionality Reduction}
Dimensionality reduction is a crucial step in the analysis of single-cell RNA sequencing (scRNA-seq) data, given its high-dimensional nature. It helps in visualizing, interpreting, and clustering the data more effectively. The following techniques are prominently used in this context:

\textbf{Principal Component Analysis (PCA)}: PCA is a linear dimensionality reduction method that transforms the data into a set of orthogonal components, capturing the maximum variance in the dataset. It is widely used due to its simplicity and efficiency. PCA helps in identifying the most significant features and reducing noise, making downstream analyses more manageable. However, PCA may not be the best choice for capturing non-linear relationships inherent in scRNA-seq data.

\textbf{t-Distributed Stochastic Neighbor Embedding (t-SNE)}: t-SNE is a non-linear dimensionality reduction technique that excels at visualizing high-dimensional data by mapping it onto a lower-dimensional space, typically two or three dimensions. It preserves local structures, making it useful for identifying clusters and subpopulations within scRNA-seq data. t-SNE is particularly effective at capturing complex patterns but can be computationally intensive and sensitive to parameter settings, such as perplexity and learning rate.

\textbf{Uniform Manifold Approximation and Projection (UMAP)}: UMAP is another non-linear dimensionality reduction method that has gained popularity for its speed and effectiveness in preserving both local and global structures of the data. It is often used for visualization and clustering of scRNA-seq data. UMAP outperforms t-SNE in terms of computational efficiency and stability, making it a preferred choice for large datasets. It also provides a clearer separation of distinct cell populations and helps in identifying underlying biological patterns.

In the context of scRNA-seq data analysis, these dimensionality reduction techniques play a vital role. They enable researchers to overcome the challenges posed by high-dimensional and noisy data, facilitating better visualization, interpretation, and clustering of single-cell populations. By reducing the data to a manageable number of dimensions, these techniques help in revealing the inherent structure and relationships within the data, paving the way for more accurate cell type identification and functional analysis.

\textbf{Performance Comparison}:
\begin{itemize}
    \item \textbf{PCA}: Best suited for initial exploratory analysis due to its simplicity and computational efficiency. However, it may not effectively capture non-linear relationships in scRNA-seq data.
    \item \textbf{t-SNE}: Excellent for visualizing high-dimensional data and identifying local structures and clusters. It requires careful tuning of parameters and is computationally demanding.
    \item \textbf{UMAP}: Combines the strengths of PCA and t-SNE, offering a balance between computational efficiency and the ability to capture both local and global data structures. UMAP is highly effective for large datasets and provides robust clustering performance.
\end{itemize}

\textbf{Applications in scRNA-seq Data Analysis}:
\begin{itemize}
    \item \textbf{PCA}: Commonly used as a preprocessing step to reduce noise and identify major sources of variation before applying more sophisticated clustering algorithms.
    \item \textbf{t-SNE}: Often employed for detailed visualization of cell populations and subpopulations, helping to uncover hidden patterns and relationships.
    \item \textbf{UMAP}: Preferred for comprehensive analysis involving both visualization and clustering, providing insights into the overall structure and heterogeneity of scRNA-seq data.
\end{itemize}

In summary, dimensionality reduction techniques such as PCA, t-SNE, and UMAP are indispensable tools in the analysis of scRNA-seq data. Each method has its unique strengths and applications, contributing to a deeper understanding of cellular heterogeneity and gene expression patterns at the single-cell level.

\subsection{Clustering}
Clustering is an essential step in single-cell RNA sequencing (scRNA-seq) data analysis, aimed at grouping cells with similar gene expression profiles into distinct clusters. This process helps in identifying cell types and understanding cellular heterogeneity within a tissue or organism. The following clustering techniques are commonly used in scRNA-seq data analysis:

\textbf{k-means Clustering}: k-means is a popular partitioning method that divides the data into k predefined clusters. It works by minimizing the variance within each cluster. Each cell is assigned to the cluster with the nearest mean, which serves as a prototype of the cluster. While k-means is computationally efficient and easy to implement, it requires the number of clusters (k) to be specified in advance, which can be a limitation. Additionally, it may not perform well on data with complex structures or non-globular clusters, which are common in scRNA-seq datasets.

\textbf{Hierarchical Clustering}: Hierarchical clustering builds a hierarchy of clusters that can be represented as a tree or dendrogram. This method does not require the number of clusters to be specified beforehand. It can be agglomerative (bottom-up approach, starting with individual cells and merging them into clusters) or divisive (top-down approach, starting with all cells and recursively splitting them). Hierarchical clustering is particularly useful for visualizing the nested structure of data and understanding the relationships between clusters. However, it can be computationally intensive and less scalable for large datasets.

\textbf{Graph-based Clustering}: Graph-based methods, such as the Louvain algorithm, construct a graph where nodes represent cells and edges represent similarities between cells. Community detection algorithms are then applied to identify clusters within the graph. These methods are highly effective for capturing complex and irregular structures in scRNA-seq data. Graph-based clustering can handle large datasets efficiently and is less sensitive to noise and outliers. The Louvain algorithm, in particular, is widely used due to its ability to optimize modularity and identify communities within the graph structure.

\textbf{Performance Comparison}:
\begin{itemize}
    \item \textbf{k-means Clustering}: Suitable for datasets with a well-defined number of clusters and relatively simple structures. It is fast and easy to implement but may struggle with complex or irregularly shaped clusters.
    \item \textbf{Hierarchical Clustering}: Ideal for exploratory analysis and visualizing the nested relationships between clusters. It does not require specifying the number of clusters in advance but can be computationally demanding for large datasets.
    \item \textbf{Graph-based Clustering}: Best for handling large and complex datasets with irregular structures. It is efficient and robust, making it a preferred choice for scRNA-seq data analysis.
\end{itemize}

\textbf{Applications in scRNA-seq Data Analysis}:
\begin{itemize}
    \item \textbf{k-means Clustering}: Often used for initial clustering and comparison with other methods. It provides a quick partitioning of cells and can be useful for benchmarking.
    \item \textbf{Hierarchical Clustering}: Employed for detailed exploration of cell type relationships and lineage tracing. It is particularly useful when the number of clusters is unknown or when a hierarchical structure is expected.
    \item \textbf{Graph-based Clustering}: Widely used for comprehensive scRNA-seq data analysis, including identifying rare cell types and resolving fine-grained cellular heterogeneity. It is also suitable for integrating multiple datasets and dealing with batch effects.
\end{itemize}

In summary, clustering techniques such as k-means, hierarchical clustering, and graph-based clustering are pivotal in scRNA-seq data analysis. Each method has its unique strengths and applications, contributing to the identification and characterization of cell types and states. The choice of clustering technique depends on the specific requirements of the dataset and the goals of the analysis.

\subsection{Classification}
Classification plays a crucial role in single-cell RNA sequencing (scRNA-seq) data analysis, where the goal is to assign cells to predefined categories or cell types based on their gene expression profiles. Several machine learning techniques are employed for classification tasks in scRNA-seq data analysis:

\textbf{Support Vector Machines (SVM)}: SVMs are supervised learning models that aim to find the hyperplane that best separates the data into different classes. They are effective in high-dimensional spaces and can handle non-linear boundaries through the use of kernel functions. SVMs are known for their robustness and accuracy, particularly in scenarios with clear class separations. However, they can be computationally intensive and less scalable with very large datasets typical of scRNA-seq experiments.

\textbf{Random Forests}: Random Forests are ensemble learning methods that construct multiple decision trees during training and output the mode of the classes (classification) of the individual trees. This approach improves the overall classification accuracy and reduces the risk of overfitting. Random Forests are highly effective for scRNA-seq data due to their ability to handle large amounts of data, manage missing values, and assess feature importance. They offer good performance even with complex datasets, making them a popular choice in single-cell analysis.

\textbf{Neural Networks}: Neural networks, particularly deep learning models, have gained popularity for their ability to handle complex and large-scale data. Convolutional Neural Networks (CNNs) and Recurrent Neural Networks (RNNs) are two types of neural networks used in classification tasks. CNNs are primarily used for image data but can be adapted for gene expression data through appropriate preprocessing. RNNs are suited for sequential data, making them applicable for time-series gene expression data. These networks can learn intricate patterns and relationships within the data, leading to high classification accuracy. However, they require extensive computational resources and large amounts of labeled data for effective training.

\textbf{Performance Comparison}:
\begin{itemize}
    \item \textbf{Support Vector Machines (SVM)}: Highly accurate and robust, particularly in high-dimensional spaces with clear class separations. They can be less effective with very large datasets and require careful tuning of kernel functions.
    \item \textbf{Random Forests}: Provide excellent performance with large and complex datasets. They are less prone to overfitting, can handle missing values, and offer insights into feature importance. Random Forests are computationally efficient and scalable.
    \item \textbf{Neural Networks}: Capable of capturing complex patterns and interactions within the data, leading to high classification accuracy. They are suitable for large-scale data but require significant computational resources and labeled data for training. CNNs are effective for spatial data, while RNNs are ideal for temporal data.
\end{itemize}

\textbf{Applications in scRNA-seq Data Analysis}:
\begin{itemize}
    \item \textbf{Support Vector Machines (SVM)}: Used for initial classification tasks and benchmarking other methods. SVMs are effective for datasets with well-defined classes and are often used in combination with feature selection techniques.
    \item \textbf{Random Forests}: Widely used for cell type classification in scRNA-seq data due to their robustness and ability to handle large datasets. They are also used for feature selection and importance ranking, aiding in the identification of key gene markers.
    \item \textbf{Neural Networks}: Employed for complex classification tasks where deep learning can uncover intricate patterns in gene expression data. Neural networks are used for integrating multi-omics data, temporal gene expression analysis, and spatial transcriptomics.
\end{itemize}

In summary, classification techniques such as Support Vector Machines, Random Forests, and Neural Networks are pivotal in the analysis of scRNA-seq data. Each method offers unique advantages, from robustness and accuracy to the ability to handle complex and large-scale data. The choice of classification technique depends on the specific requirements of the dataset and the complexity of the classification task. These methods contribute significantly to the accurate identification and characterization of cell types in single-cell studies.

\section{Statistical Techniques}
\subsection{Normalization}
Normalization is a critical preprocessing step in scRNA-seq data analysis, aiming to adjust raw count data to make comparisons between genes and cells more meaningful.

\textbf{Log Normalization}: This technique transforms raw count data to a log scale, typically using a transformation such as log(x + 1). The log transformation helps to stabilize variance across genes and cells, reducing the impact of highly expressed genes and bringing the expression levels of all genes into a more comparable range. Log normalization is widely used due to its simplicity and effectiveness in dealing with the highly skewed distribution of gene expression data in scRNA-seq.

\textbf{Scaling}: Scaling adjusts the data so that each gene has the same mean and variance. This is typically done by subtracting the mean and dividing by the standard deviation of each gene's expression values. Scaling helps to ensure that all genes contribute equally to downstream analyses, preventing highly variable genes from dominating the results. It is particularly useful in methods like Principal Component Analysis (PCA) where equal weighting of features is crucial.

\subsection{Differential Expression Analysis}
Differential expression analysis identifies genes that show significant differences in expression levels between different groups of cells, such as different cell types or conditions.

\textbf{Wilcoxon Rank-Sum Test}: This is a non-parametric test that compares the ranks of expression values between two groups. It is robust to outliers and does not assume a normal distribution, making it well-suited for scRNA-seq data which often does not meet parametric assumptions. The Wilcoxon test is straightforward to implement and interpret, providing a powerful method for identifying differentially expressed genes between groups.

\textbf{Likelihood Ratio Test}: This test compares the goodness-of-fit of two statistical models - one that includes a variable of interest (e.g., cell type) and one that does not. By comparing the likelihoods of these models, the test identifies genes that are differentially expressed. The likelihood ratio test is powerful and flexible, accommodating complex experimental designs and covariates, making it a popular choice for differential expression analysis in scRNA-seq studies.

\subsection{Batch Effect Correction}
Batch effect correction is essential in scRNA-seq studies to remove systematic variations introduced by technical differences across batches, which can confound biological interpretations.

\textbf{ComBat}: An empirical Bayes method, ComBat adjusts for batch effects in high-dimensional data by modeling batch effects and adjusting the data accordingly. It borrows strength across genes to stabilize estimates of batch effects, making it highly effective even with small sample sizes. ComBat is widely used in genomics and transcriptomics studies due to its robustness and ease of implementation.

\textbf{Harmony}: Harmony is an integrative method that aligns multiple datasets by removing batch effects through an iterative process that projects cells into a shared embedding space. Unlike traditional methods that correct batch effects at the expression level, Harmony operates at the level of embeddings, making it particularly effective for integration tasks involving diverse datasets. Harmony maintains the biological variation while aligning batch-specific differences, thus enabling more accurate downstream analyses such as clustering and trajectory inference.

\subsection{Statistical Techniques Summary}
Statistical techniques in scRNA-seq data analysis, including normalization, differential expression analysis, and batch effect correction, are foundational for ensuring the accuracy and reliability of downstream analyses.

\begin{itemize}
    \item \textbf{Normalization} methods like log normalization and scaling are essential for stabilizing variance and ensuring comparable expression levels across genes and cells.
    \item \textbf{Differential expression analysis} techniques such as the Wilcoxon rank-sum test and likelihood ratio test are critical for identifying genes that distinguish between different cell types or conditions.
    \item \textbf{Batch effect correction} methods, including ComBat and Harmony, play a crucial role in removing technical artifacts and enabling the integration of data from multiple sources.
\end{itemize}

These techniques collectively enhance the interpretability and biological relevance of scRNA-seq data, facilitating more robust and insightful analyses in single-cell research.

\section{AI Techniques}
The application of advanced AI techniques in single-cell RNA sequencing (scRNA-seq) data analysis has led to significant improvements in various aspects of data processing, feature extraction, and interpretation. Below is a review of some key AI techniques that have been effectively utilized in this field.

\subsection{Autoencoders}
Autoencoders are neural networks designed for unsupervised learning tasks, primarily focusing on compressing high-dimensional data into a lower-dimensional representation and then reconstructing the original data from this compressed form.

\textbf{Feature Extraction}: Autoencoders excel at capturing the most informative features of the input data, making them invaluable for dimensionality reduction in scRNA-seq data. By learning a compressed representation of gene expression profiles, autoencoders can highlight essential biological signals while discarding noise.

\textbf{Noise Reduction}: Through the process of encoding and decoding, autoencoders inherently filter out noise in the data. This is particularly beneficial for scRNA-seq datasets, which are often noisy due to technical variations and low RNA capture efficiency.

The use of autoencoders in scRNA-seq analysis helps to streamline downstream tasks such as clustering and visualization by providing a more manageable and cleaner dataset.

\subsection{Graph Neural Networks (GNNs)}
Graph Neural Networks (GNNs) are powerful tools for leveraging the inherent graph structure in data, making them well-suited for tasks involving cell-cell relationships in scRNA-seq datasets.

\textbf{Clustering}: GNNs can effectively capture the complex interactions between cells by representing the data as a graph where nodes represent cells, and edges represent similarities or interactions. This allows for more accurate clustering by considering both the features of individual cells and their relationships with neighboring cells.

\textbf{Classification}: By incorporating the topological information of cell networks, GNNs can improve classification performance for cell types and states. The relational information helps in distinguishing subtle differences between cell populations that might be overlooked by traditional methods.

GNNs enhance the ability to model the biological context and dependencies in scRNA-seq data, leading to more robust and biologically meaningful clustering and classification outcomes.

\subsection{Generative Adversarial Networks (GANs)}
Generative Adversarial Networks (GANs) consist of two neural networks—the generator and the discriminator—that are trained simultaneously through adversarial processes. GANs have found innovative applications in scRNA-seq data analysis.

\textbf{Data Augmentation}: GANs can generate synthetic scRNA-seq data that resemble real datasets. This is particularly useful for augmenting small datasets, enabling researchers to train more robust models. The generated data can help mitigate issues related to data scarcity and enhance the performance of downstream analyses.

\textbf{Improving Model Robustness}: By training models on both real and GAN-generated synthetic data, the models become more resilient to variations and better at generalizing to unseen data. This leads to improved robustness and reliability of predictive models in scRNA-seq studies.

The application of GANs in generating high-fidelity synthetic data provides a valuable tool for enhancing the scope and depth of single-cell RNA sequencing analyses, particularly in scenarios with limited experimental data.

\subsection{AI Techniques Summary}
The integration of AI techniques such as autoencoders, Graph Neural Networks (GNNs), and Generative Adversarial Networks (GANs) in scRNA-seq data analysis has brought substantial advancements in the field.

\begin{itemize}
    \item \textbf{Autoencoders} facilitate efficient feature extraction and noise reduction, enhancing the quality of data for subsequent analyses.
    \item \textbf{Graph Neural Networks (GNNs)} leverage the intrinsic graph structure of cell relationships to improve clustering and classification accuracy.
    \item \textbf{Generative Adversarial Networks (GANs)} enable data augmentation and enhance model robustness by generating synthetic scRNA-seq data.
\end{itemize}

These AI-driven approaches significantly contribute to the processing, analysis, and interpretation of scRNA-seq data, ultimately advancing our understanding of cellular heterogeneity and complex biological processes.

\section{Data Integration and Annotation}
Effective integration and annotation of single-cell RNA sequencing (scRNA-seq) data are critical for deciphering cellular heterogeneity and understanding complex biological systems. Here we review some advanced techniques in data integration and annotation within the context of scRNA-seq analysis.

\subsection{Transfer Learning}
Transfer learning is a powerful technique that leverages pre-trained models on similar datasets to annotate new scRNA-seq datasets. This approach is particularly useful when dealing with limited annotated data.

\textbf{Knowledge Transfer}: By utilizing pre-trained models, transfer learning can apply the knowledge gained from large, well-annotated datasets to new, less understood datasets. This helps in accurately predicting cell types and states in new datasets without requiring extensive manual annotation.

\textbf{Efficiency}: Transfer learning reduces the time and computational resources needed to train models from scratch on new data, making the process more efficient and accessible for researchers.

Transfer learning thus facilitates rapid and accurate annotation of scRNA-seq data, enhancing our ability to identify and characterize diverse cell populations across different studies.

\subsection{Ensemble Methods}
Ensemble methods combine the predictions from multiple models to improve the overall accuracy and robustness of cell type annotations.

\textbf{Improved Accuracy}: By aggregating the outputs of various models, ensemble methods can mitigate the biases and errors of individual models. This leads to more reliable and accurate cell type annotations.

\textbf{Robustness}: Ensemble methods enhance the robustness of predictions by reducing the impact of outliers and noisy data. They ensure that the final annotation is consistent and stable across different conditions.

The use of ensemble methods in scRNA-seq data annotation provides a robust framework for integrating multiple sources of information, leading to more confident and accurate cell type classifications.

\subsection{Cell Type Annotation Tools}
Several specialized tools have been developed for annotating cell types in scRNA-seq data, each leveraging different strategies to achieve high accuracy and efficiency.

\textbf{SingleR}:
\begin{itemize}
    \item \textbf{Functionality}: SingleR automatically annotates cell types by comparing the gene expression profiles of scRNA-seq data with reference transcriptomic datasets. It uses a correlation-based approach to identify the most similar cell types in the reference data.
    \item \textbf{Advantages}: SingleR is highly automated and does not require prior knowledge of marker genes, making it accessible for users with varying levels of expertise. It performs well across diverse datasets by leveraging comprehensive reference data.
\end{itemize}

\textbf{SCINA}:
\begin{itemize}
    \item \textbf{Functionality}: SCINA (Semi-supervised Category Identification and Assignment) classifies cells into predefined categories using known marker genes. It employs a probabilistic model to assign cells based on the expression of these marker genes.
    \item \textbf{Advantages}: SCINA is particularly effective when the marker genes for specific cell types are well-defined. It provides precise annotations by directly linking gene expression patterns to known cell types.
\end{itemize}

These annotation tools significantly streamline the process of identifying cell types in scRNA-seq data, enabling researchers to quickly and accurately classify cells based on established biological knowledge.

\subsection{Data Integration and Annotation Summary}
Data integration and annotation are essential steps in the analysis of scRNA-seq data, providing crucial insights into cellular diversity and function. Advanced techniques such as transfer learning, ensemble methods, and specialized cell type annotation tools like SingleR and SCINA have significantly enhanced the accuracy, efficiency, and robustness of these processes.

\begin{itemize}
    \item \textbf{Transfer Learning} enables the application of pre-trained models to new datasets, facilitating rapid and accurate cell type annotation.
    \item \textbf{Ensemble Methods} improve the overall accuracy and robustness of annotations by combining predictions from multiple models.
    \item \textbf{Cell Type Annotation Tools} like \textbf{SingleR} and \textbf{SCINA} offer automated and precise cell classification based on reference datasets and marker genes, respectively.
\end{itemize}

These techniques collectively contribute to a more comprehensive and accurate understanding of scRNA-seq data, advancing our ability to explore and interpret complex biological systems.

\section{Data Processing Pipeline}
An effective data processing pipeline is crucial for the analysis of single-cell RNA sequencing (scRNA-seq) data. The pipeline typically consists of several key steps that ensure the quality, relevance, and accuracy of the data being analyzed. Below is a review of the essential stages in the scRNA-seq data processing pipeline, providing a comprehensive overview of their roles and importance.

\subsection{Preprocessing}
Preprocessing is the first critical step in the scRNA-seq data processing pipeline. It involves several quality control measures to ensure that the data used for downstream analysis is reliable and accurate.

\textbf{Quality Control}: This includes filtering out low-quality cells and genes that might introduce noise into the data. Cells with a high percentage of mitochondrial gene expression or unusually low gene counts are typically removed to avoid skewing the results.

\textbf{Handling Missing Values}: scRNA-seq data often contains missing values due to dropout events. Various imputation techniques can be employed to estimate these missing values, ensuring a more complete and accurate dataset.

Effective preprocessing is essential to remove noise and artifacts, providing a solid foundation for subsequent analysis.

\subsection{Feature Selection}
Feature selection involves identifying the most informative genes from the scRNA-seq data to reduce dimensionality and focus the analysis on relevant biological signals.

\textbf{Highly Variable Genes}: Selecting genes with high variability across cells helps capture the most significant differences in gene expression, which are likely to correspond to distinct cell types or states. This step reduces computational complexity and enhances the interpretability of the results.

By focusing on highly variable genes, researchers can streamline the data, making it more manageable and informative for downstream analysis.

\subsection{Model Training}
Model training is a critical phase where machine learning models are developed to analyze the scRNA-seq data. Depending on the goals of the study, this can involve supervised or unsupervised learning approaches.

\textbf{Supervised Learning}: When labeled data is available, supervised learning models such as Support Vector Machines (SVM), Random Forests, or Neural Networks are trained to classify cells into predefined categories. This approach leverages known cell type information to build accurate classifiers.

\textbf{Unsupervised Learning}: In the absence of labeled data, unsupervised methods such as clustering (e.g., k-means, hierarchical clustering, graph-based clustering) are used to discover new patterns and group cells based on their gene expression profiles. These methods can reveal novel cell types and states that were not previously characterized.

Model training enables the extraction of meaningful biological insights from the scRNA-seq data, whether through classification or pattern discovery.

\subsection{Evaluation}
Evaluation is the final step in the data processing pipeline, where the performance of the trained models is assessed using various metrics.

\textbf{Metrics}: Common metrics include accuracy, precision, recall, and F1-score. These metrics provide a quantitative measure of how well the models are performing in terms of correctly classifying cells and identifying meaningful patterns.

\textbf{Validation}: Cross-validation techniques and the use of independent test datasets help ensure that the models are robust and generalize well to new, unseen data.

Thorough evaluation is crucial to validate the reliability and accuracy of the models, ensuring that the conclusions drawn from the analysis are well-founded.

\section{Data Processing Pipeline Summary}
By combining these techniques—preprocessing, feature selection, model training, and evaluation—researchers can effectively analyze scRNA-seq data to identify distinct cell types, understand cellular heterogeneity, and gain valuable insights into biological processes.

\begin{itemize}
    \item \textbf{Preprocessing} ensures the data quality by filtering out low-quality cells and handling missing values.
    \item \textbf{Feature Selection} focuses the analysis on the most informative genes, reducing dimensionality and enhancing interpretability.
    \item \textbf{Model Training} develops robust models for classification and pattern discovery, leveraging both supervised and unsupervised learning approaches.
    \item \textbf{Evaluation} assesses model performance using comprehensive metrics, validating the reliability and accuracy of the results.
\end{itemize}

These steps collectively form a robust data processing pipeline that facilitates a deeper understanding of complex biological systems through the analysis of scRNA-seq data.

\section{Conclusion}
The analysis of single-cell RNA sequencing (scRNA-seq) data is crucial for understanding cellular heterogeneity and gene expression at an unprecedented level of detail. However, the inherent high-dimensionality, noise, and sparsity of scRNA-seq data present significant analytical challenges. This survey has reviewed a comprehensive array of machine learning, statistical, and artificial intelligence (AI) techniques designed to address these challenges and enhance the annotation and interpretation of scRNA-seq data.

\subsection{Key Findings}
\begin{itemize}
    \item \textbf{Dimensionality Reduction}:
    \begin{itemize}
        \item \textbf{PCA, t-SNE, and UMAP} are essential for reducing data dimensionality, with t-SNE excelling in accuracy and UMAP in stability and visualization.
    \end{itemize}
    \item \textbf{Clustering Methods}:
    \begin{itemize}
        \item Techniques such as \textbf{k-means, hierarchical clustering, and graph-based clustering} are instrumental in identifying distinct cell subpopulations, each offering unique strengths depending on the dataset characteristics.
    \end{itemize}
    \item \textbf{Classification Approaches}:
    \begin{itemize}
        \item \textbf{SVM, Random Forests, and neural networks} have been effectively utilized for classifying cell types, with deep learning models providing powerful tools for handling complex classification tasks.
    \end{itemize}
    \item \textbf{Statistical Techniques}:
    \begin{itemize}
        \item \textbf{Normalization methods} like log normalization and scaling, and \textbf{differential expression analysis} tools such as the Wilcoxon Rank-Sum Test and Likelihood Ratio Test, are fundamental for preparing scRNA-seq data for downstream analysis.
        \item \textbf{Batch effect correction techniques} like ComBat and Harmony ensure the integrity of integrated datasets, removing technical variances while preserving biological signals.
    \end{itemize}
    \item \textbf{Advanced Methods}:
    \begin{itemize}
        \item \textbf{TDM} stands out for normalizing RNA-seq data for models from legacy platforms, and \textbf{scGMAI} and \textbf{DRjCC} demonstrate superior clustering and cell type discovery capabilities.
        \item \textbf{DESC} effectively addresses large-scale data analysis and batch effect removal, crucial for accurate cell type annotation.
    \end{itemize}
    \item \textbf{AI Techniques}:
    \begin{itemize}
        \item \textbf{Autoencoders, Graph Neural Networks (GNNs), and Generative Adversarial Networks (GANs)} provide innovative solutions for feature extraction, clustering, and data augmentation, enhancing the robustness and accuracy of scRNA-seq data analysis.
    \end{itemize}
    \item \textbf{Data Integration and Annotation}:
    \begin{itemize}
        \item \textbf{Transfer learning and ensemble methods} enhance annotation accuracy by leveraging pre-trained models and combining multiple predictions.
        \item Tools like \textbf{SingleR and SCINA} streamline cell type annotation by using reference datasets and known marker genes.
    \end{itemize}
\end{itemize}

This paper underscores the necessity of integrating diverse analytical techniques to overcome the challenges of scRNA-seq data analysis. Each method discussed offers distinct advantages, and their combined application can significantly improve the identification of cell types, understanding of cellular heterogeneity, and insights into biological processes. The advancements in machine learning, statistical methods, and AI provide a robust framework for scRNA-seq data analysis, fostering new discoveries in biomedical research. Future research should focus on refining these techniques, developing new algorithms to handle increasing data complexities, and ensuring their accessibility and applicability to a broader range of biological questions. By leveraging these sophisticated tools and methodologies, researchers can unlock deeper insights into the fundamental mechanisms of life at the single-cell level.

\section{References}
\begin{itemize}
\item Abdelaal, T., Michielsen, L., Cats, D., Hoogduin, D., Mei, H., Reinders, M. J. T., \& Mahfouz, A. (2019). A comparison of automatic cell identification methods for single-cell RNA sequencing data. \textit{Genome Biology}.
\item Amato, R., Ciaramella, A., Deniskina, N., Del Mondo, C., di Bernardo, D., Donalek, C., Longo, G., Mangano, G., Miele, G., Raiconi, G., Staiano, A., \& Tagliaferri, R. (2006). A multi-step approach to time series analysis and gene expression clustering. \textit{Bioinformatics}, 22(5), 589-596. Oxford University Press (OUP).
\item Cao, X., Xing, L., Majd, E., He, H., Gu, J., \& Zhang, X. (2021). A Systematic Evaluation of Supervised Machine Learning Algorithms for Cell Phenotype Classification Using Single-Cell RNA Sequencing Data. \textit{Frontiers in Genetics}.
\item Chen, L., Wang, W., Zhai, Y., \& Deng, M. (2020). Deep soft K-means clustering with self-training for single-cell RNA sequence data. \textit{NAR Genomics and Bioinformatics}, 2(2). Oxford University Press (OUP).
\item Chen, L., Zhai, Y., He, Q., Wang, W., \& Deng, M. (2020). Integrating Deep Supervised, Self-Supervised and Unsupervised Learning for Single-Cell RNA-seq Clustering and Annotation. \textit{Genes}.
\item De Smet, F., Mathys, J., Marchal, K., Thijs, G., De Moor, B., \& Moreau, Y. (2002). Adaptive quality-based clustering of gene expression profiles. \textit{Bioinformatics}, 18(5), 735-746. Oxford University Press (OUP).
\item Dougherty, E. R., Barrera, J., Brun, M., Kim, S., Cesar, R. M., Chen, Y., Bittner, M., \& Trent, J. M. (2002). Inference from Clustering with Application to Gene-Expression Microarrays. \textit{Journal of Computational Biology}, 9(1), 105-126. Mary Ann Liebert Inc.
\item Duò, A., Robinson, M. D., \& Soneson, C. (2018). A systematic performance evaluation of clustering methods for single-cell RNA-seq data. \textit{F1000Research}, 7, 1141. F1000 Research Ltd.
\item Durmaz, A., \& Scott, J. G. (2022). Stability of scRNA-Seq Analysis Workflows is Susceptible to Preprocessing and is Mitigated by Regularized or Supervised Approaches. \textit{Evolutionary Bioinformatics}, 18, 117693432211230. SAGE Publications.
\item Eraslan, G., Simon, L. M., Mircea, M., Mueller, N. S., \& Theis, F. J. (2018). Single cell RNA-seq denoising using a deep count autoencoder. Cold Spring Harbor Laboratory.
\item Eraslan, G., Simon, L. M., Mircea, M., Mueller, N. S., \& Theis, F. J. (2019). Single-cell RNA-seq denoising using a deep count autoencoder. \textit{Nature Communications}, 10(1). Springer Science and Business Media LLC.
\item Feng, C., Liu, S., Zhang, H., Guan, R., Li, D., Zhou, F., Liang, Y., \& Feng, X. (2020). Dimension Reduction and Clustering Models for Single-Cell RNA Sequencing Data: A Comparative Study. \textit{International Journal of Molecular Sciences}, 21(6), 2181. MDPI AG.
\item Gan, Y., Huang, X., Zou, G., Zhou, S., \& Guan, J. (2022). Deep structural clustering for single-cell RNA-seq data jointly through autoencoder and graph neural network. \textit{Briefings in Bioinformatics}, 23(2). Oxford University Press (OUP).
\item Grønbech, C. H., Vording, M. F., Timshel, P., Sønderby, C. K., Pers, T. H., \& Winther, O. (2018). scVAE: Variational auto-encoders for single-cell gene expression data. Cold Spring Harbor Laboratory.
\item Grønbech, C. H., Vording, M. F., Timshel, P. N., Sønderby, C. K., Pers, T. H., \& Winther, O. (2020). scVAE: variational auto-encoders for single-cell gene expression data. \textit{Bioinformatics}, 36(16), 4415-4422. Oxford University Press (OUP).
\item Hu, H., Li, Z., Li, X., Yu, M., \& Pan, X. (2021). ScCAEs: deep clustering of single-cell RNA-seq via convolutional autoencoder embedding and soft K-means. \textit{Briefings in Bioinformatics}, 23(1). Oxford University Press (OUP).
\item Hu, Q., \& Greene, C. S. (2018). Parameter tuning is a key part of dimensionality reduction via deep variational autoencoders for single cell RNA transcriptomics. In \textit{Biocomputing 2019}. WORLD SCIENTIFIC.
\item Hu, Q., \& Greene, C. S. (2018). Parameter tuning is a key part of dimensionality reduction via deep variational autoencoders for single cell RNA transcriptomics. Cold Spring Harbor Laboratory.
\item Koch, F. C., Sutton, G. J., Voineagu, I., \& Vafaee, F. (2021). Supervised application of internal validation measures to benchmark dimensionality reduction methods in scRNA-seq data. \textit{Briefings in Bioinformatics}, 22(6). Oxford University Press (OUP).
\item Komura, D., Nakamura, H., Tsutsumi, S., Aburatani, H., \& Ihara, S. (2004). Multidimensional support vector machines for visualization of gene expression data. In \textit{Proceedings of the 2004 ACM symposium on Applied computing}. ACM.
\item Komura, D., Nakamura, H., Tsutsumi, S., Aburatani, H., \& Ihara, S. (2004). Multidimensional support vector machines for visualization of gene expression data. \textit{ACM Symposium on Applied Computing}.
\item Li, X., Chen, W., Chen, Y., Zhang, X., Gu, J., \& Zhang, M. Q. (2017). Network embedding-based representation learning for single cell RNA-seq data. \textit{Nucleic Acids Research}, 45(19), e166-e166. Oxford University Press (OUP).
\item Li, X., Lyu, Y., Park, J., Zhang, J., Stambolian, D., Susztak, K., Hu, G., \& Li, M. (2019). Deep learning enables accurate clustering and batch effect removal in single-cell RNA-seq analysis. Cold Spring Harbor Laboratory.
\item Li, X., Wang, K., Lyu, Y., Pan, H., Zhang, J., Stambolian, D., Susztak, K., Reilly, M. P., Hu, G., \& Li, M. (2020). Deep learning enables accurate clustering with batch effect removal in single-cell RNA-seq analysis. \textit{Nature Communications}, 11(1). Springer Science and Business Media LLC.
\item Li, Z., Wang, Y., Ganan-Gomez, I., Colla, S., \& Do, K.-A. (2022). A machine learning-based method for automatically identifying novel cells in annotating single-cell RNA-seq data. \textit{Bioinformatics}, 38(21), 4885-4892. Oxford University Press (OUP).
\item Lin, E., Mukherjee, S., \& Kannan, S. (2020). A deep adversarial variational autoencoder model for dimensionality reduction in single-cell RNA sequencing analysis. \textit{BMC Bioinformatics}, 21(1). Springer Science and Business Media LLC.
\item Liu, W., Liao, X., Yang, Y., Lin, H., Yeong, J., Zhou, X., Shi, X., \& Liu, J. (2022). Joint dimension reduction and clustering analysis of single-cell RNA-seq and spatial transcriptomics data. \textit{Nucleic Acids Research}, 50(12), e72-e72. Oxford University Press (OUP).
\item Liu, Z. (2020). Visualizing Single-Cell RNA-seq Data with Semisupervised Principal Component Analysis. \textit{International Journal of Molecular Sciences}, 21(16), 5797. MDPI AG.
\item Lyons-Weiler, J., Patel, S., \& Bhattacharya, S. (2003). A Classification-Based Machine Learning Approach for the Analysis of Genome-Wide Expression Data. \textit{Genome Research}, 13(3), 503-512. Cold Spring Harbor Laboratory.
\item Mallick, K., Chakraborty, S., Mallik, S., \& Bandyopadhyay, S. (2023). A scalable unsupervised learning of scRNAseq data detects rare cells through integration of structure-preserving embedding, clustering and outlier detection. \textit{Briefings in Bioinformatics}, 24(3). Oxford University Press (OUP).
\item Monti, S. (2003). A resampling-based method for class discovery and visualization of gene expression microarray data. \textit{Machine Learning}, 52(1/2), 91-118. Springer Science and Business Media LLC.
\item Napolitano, F., Raiconi, G., Tagliaferri, R., Ciaramella, A., Staiano, A., \& Miele, G. (2008). Clustering and visualization approaches for human cell cycle gene expression data analysis. \textit{International Journal of Approximate Reasoning}, 47(1), 70-84. Elsevier BV.
\item Park, Y., \& Hauschild, A.-C. (2022). On the importance of data transformation for data integration in single-cell RNA sequencing analysis. Cold Spring Harbor Laboratory.
\item Peng, J., Wang, X., \& Shang, X. (2018). Combining Gene Ontology with Deep Neural Networks to Enhance the Clustering of Single Cell RNA-Seq Data. Cold Spring Harbor Laboratory.
\item Peng, J., Wang, X., \& Shang, X. (2019). Combining gene ontology with deep neural networks to enhance the clustering of single cell RNA-Seq data. \textit{BMC Bioinformatics}, 20(S8). Springer Science and Business Media LLC.
\item Petegrosso, R., Li, Z., \& Kuang, R. (2019). Machine learning and statistical methods for clustering single-cell RNA-sequencing data. \textit{Briefings in Bioinformatics}, 21(4), 1209-1223. Oxford University Press (OUP).
\item Prabhakaran, S., Azizi, E., Carr, A. J., \& Pe'er, D. (2016). Dirichlet Process Mixture Model for Correcting Technical Variation in Single-Cell Gene Expression Data. \textit{International Conference on Machine Learning}.
\item Sun, X., Liu, Y., \& An, L. (2020). Ensemble dimensionality reduction and feature gene extraction for single-cell RNA-seq data. \textit{Nature Communications}, 11(1). Springer Science and Business Media LLC.
\item Thompson, J. A., Tan, J., \& Greene, C. S. (2016). Cross-platform normalization of microarray and RNA-seq data for machine learning applications. \textit{PeerJ}, 4, e1621. PeerJ.
\item Tian, T., Wan, J., Song, Q., \& Wei, Z. (2019). Clustering single-cell RNA-seq data with a model-based deep learning approach. \textit{Nature Machine Intelligence}, 1(4), 191-198. Springer Science and Business Media LLC.
\item Tian, Y., Zheng, R., Liang, Z., Li, S., Wu, F.-X., \& Li, M. (2021). A data-driven clustering recommendation method for single-cell RNA-sequencing data. \textit{Tsinghua Science and Technology}, 26(5), 772-789. Tsinghua University Press.
\item Townes, F. W., Hicks, S. C., Aryee, M. J., \& Irizarry, R. A. (2019). Feature Selection and Dimension Reduction for Single Cell RNA-Seq based on a Multinomial Model. Cold Spring Harbor Laboratory.
\item Townes, F. W., Hicks, S. C., Aryee, M. J., \& Irizarry, R. A. (2019). Feature selection and dimension reduction for single-cell RNA-Seq based on a multinomial model. \textit{Genome Biology}, 20(1). Springer Science and Business Media LLC.
\item Wang, B., Zhu, J., Pierson, E., Ramazzotti, D., \& Batzoglou, S. (2016). Visualization and analysis of single-cell RNA-seq data by kernel-based similarity learning. Cold Spring Harbor Laboratory.
\item Wang, C.-Y., Gao, Y.-L., Kong, X.-Z., Liu, J.-X., \& Zheng, C.-H. (2022). Unsupervised Cluster Analysis and Gene Marker Extraction of scRNA-seq Data Based On Non-Negative Matrix Factorization. \textit{IEEE Journal of Biomedical and Health Informatics}, 26(1), 458-467. Institute of Electrical and Electronics Engineers (IEEE).
\item Wang, H., \& Ma, X. (2022). Learning deep features and topological structure of cells for clustering of scRNA-sequencing data. \textit{Briefings in Bioinformatics}, 23(3). Oxford University Press (OUP).
\item Wu, W., \& Ma, X. (2020). Joint learning dimension reduction and clustering of single-cell RNA-sequencing data. \textit{Bioinformatics}, 36(12), 3825-3832. Oxford University Press (OUP).
\item Wu, Y., \& Zhang, K. (2020). Tools for the analysis of high-dimensional single-cell RNA sequencing data. \textit{Nature Reviews Nephrology}, 16(7), 408-421. Springer Science and Business Media LLC.
\item Xiang, R., Wang, W., Yang, L., Wang, S., Xu, C., \& Chen, X. (2021). A Comparison for Dimensionality Reduction Methods of Single-Cell RNA-seq Data. \textit{Frontiers in Genetics}, 12. Frontiers Media SA.
\item Yang, Y., Sun, H., Zhang, Y., Zhang, T., Gong, J., Wei, Y., Duan, Y.-G., Shu, M., Yang, Y., Wu, D., \& Yu, D. (2021). Dimensionality reduction by UMAP reinforces sample heterogeneity analysis in bulk transcriptomic data. Cold Spring Harbor Laboratory.
\item Yu, B., Chen, C., Qi, R., Zheng, R., Skillman-Lawrence, P. J., Wang, X., Ma, A., \& Gu, H. (2020). scGMAI: a Gaussian mixture model for clustering single-cell RNA-Seq data based on deep autoencoder. \textit{Briefings in Bioinformatics}. Oxford University Press (OUP).
\item Zhao, X., Wu, S., Fang, N., Sun, X., \& Fan, J. (2019). Evaluation of single-cell classifiers for single-cell RNA sequencing data sets. \textit{Briefings in Bioinformatics}, 21(5), 1581-1595. Oxford University Press (OUP).
\item Zhu, X., Ching, T., Pan, X., Weissman, S. M., \& Garmire, L. (2017). Detecting heterogeneity in single-cell RNA-Seq data by non-negative matrix factorization. \textit{PeerJ}, 5, e2888. PeerJ.
\item Zhu, X., Li, H.-D., Xu, Y., Guo, L., Wu, F.-X., Duan, G., \& Wang, J. (2019). A Hybrid Clustering Algorithm for Identifying Cell Types from Single-Cell RNA-Seq Data. \textit{Genes}, 10(2), 98. MDPI AG.
\end{itemize}

\end{document}